\documentclass[11pt]{article}
\usepackage{mainstyle}
\usepackage[utf8]{inputenc}
\usepackage[noend]{algpseudocode}
\usepackage[backend=bibtex, maxnames=4, sorting=none, citestyle=numeric-comp, bibstyle=trad-unsrt]{biblatex}
\usepackage{soul}

\usepackage[a4paper,
            left=2cm,
            right=2cm,
            top=2cm,
            bottom=2cm]{geometry}

\title{Effects of near-surface sedimentary structure on Newtonian noise for the Einstein Telescope: a 2-D numerical study}

\author{Shi Yao$^{1,2\,*}$, Patrick Schillings $^{3\,*}$, Johannes Erdmann$^3$, Andreas Rietbrock$^1$}

\date{\small$^1$ Karlsruhe Institute of Technology, Geophysical Institute, Karlsruhe, Germany\\
\small$^2$ Technische Universität Dresden, Institute for Nuclear and Particle Physics (IKTP), Dresden, Germany\\
\small$^3$ RWTH Aachen University, III. Physikalisches Institut A, Aachen, Germany\\}

\begin{document}

\maketitle

\begingroup
\renewcommand{\thefootnote}{\fnsymbol{footnote}}
\footnotetext[1]{Both authors contributed equally.}
\footnotetext[2]{\texttt{shi.yao@tu-dresden.de}}
\footnotetext[3]{\texttt{patrick.schillings@rwth-aachen.de}}
\endgroup

\begin{abstract}
\noindent
Near-surface low-velocity sediments can strongly modify seismic wavefields and therefore affect estimates of Newtonian noise at underground gravitational-wave observatories. We investigate these effects using 2-D viscoelastic simulations of a sediment layer overlying hard-rock basement. Controlled experiments examine the influence of sediment-basement interface geometry, sediment shear-wave velocity, attenuation, and test-mass position. Relative to a homogeneous model, the sediment layer produces frequency-dependent changes through wave trapping, interference, and attenuation. A constant-thickness layer generates a more laterally coherent wavefield and a sharper spectral enhancement than a basin-shaped interface, whereas lateral thickness variations broaden and shift the response. Sediment shear-wave velocity primarily controls the frequencies of the spectral features, while attenuation mainly controls their amplitudes. Newtonian noise is most sensitive to sediment structure for test masses located within or several hundred meters below the sediment layer. The sensitivity to lateral variations in sediment thickness decreases with burial depth and is weak at $2~\mathrm{km}$ depth in the present model. These results demonstrate that near-surface sedimentary structure should be represented explicitly when assessing site-dependent Newtonian noise, particularly for test masses located between $200~\mathrm{m}$ and $300~\mathrm{m}$ depths. \\

\noindent
Keywords: Newtonian noise, Numerical simulation, Sediment layer, Einstein Telescope
\end{abstract}

\section{Introduction}

The Einstein Telescope (ET) is a proposed third-generation underground gravitational-wave observatory designed to extend the sensitivity of ground-based detectors to frequencies of a few hertz \cite{ETDesignReportUpdate}. Access to this low-frequency band would broaden the observable frequency range of gravitational-wave signals, but it also requires strict conditions regarding the seismic environment and subsurface conditions at the detector site \cite{amann2020site, lowerLimit}. In particular, seismic displacement produces time-dependent density perturbations in the surrounding medium, which generate fluctuating gravitational forces on the suspended test masses \cite{EarlyLIGO_NN, EarlyVirgo_NN, terrestialGravityFluctuations, NewtonianNoiseOrigin}. This contribution, known as Newtonian noise, cannot be eliminated by mechanical isolation and is expected to limit the detector sensitivity at low frequencies \cite{HildsNoiseEstimation}. Reliable site evaluation therefore requires not only measurements of ambient seismic noise, but also a detailed understanding of how the local geological structure modifies the seismic wavefield and the resulting Newtonian noise.

Current site-characterization studies demonstrate that near-surface geological structure is an important consideration at all three candidate ET sites. At the Euregio Meuse-Rhine site (EMR) candidate site, hard Paleozoic basement rocks are overlain by soft, partly unconsolidated Cretaceous and Cenozoic sediments, producing strong vertical and lateral contrasts in near-surface material properties \cite{burchartz2025emr}. The Sos Enattos region in Sardinia is dominated by exposed metamorphic and granitoid basement rather than a thick sedimentary cover. However, recent studies using borehole data, electrical resistivity imaging, and seismic tomography have identified a weathered near-surface layer and localized fractured zones with reduced seismic velocities \cite{villani2025subsurface}, while Rayleigh-wave dispersion inversion resolves a depth-dependent near-surface shear-wave velocity structure \cite{saccorotti2023array}. In the Lusatia, ongoing investigations combine dense seismic arrays, active-source profiles, borehole measurements, and geological modelling to characterize the near-surface shear-wave velocity structure and the sedimentary cover above the granodioritic basement \cite{ryberg2026lausitz}. These observations indicate that near-surface low-velocity layers and sharp material contrasts occur in different geological forms across the candidate sites and must therefore be considered when modelling Newtonian noise.

Low-velocity sedimentary layers are well known to produce strong, frequency-dependent effects. The impedance contrast across the sediment-basement interface causes repeated reflections and traps seismic energy within the sediment, leading to amplified motion at characteristic frequencies controlled primarily by the sediment thickness and shear-wave velocity \cite{bard1985twodimensional,roten2006twodimensional}. For laterally uniform layers, the response can often be approximated as 1-D resonance. However, in sedimentary basins, lateral variations in interface geometry generate additional wave conversion, diffraction, basin-edge-generated surface waves, and spatially variable interference, resulting in spatially variable amplification and more complex wavefields \cite{semblat2005basin,tian2025basinanalysis}. Observations from dense seismic arrays also demonstrate that site amplification, intrinsic attenuation, and scattering can make distinct contributions to the recorded wavefield \cite{bowden2015site}. These studies indicate that sediment thickness, shear-wave velocity, attenuation, and interface geometry jointly control the amplitude, frequency content, and spatial coherence of near-surface seismic motion.

Despite the well-established influence of sedimentary layers on seismic ground motion, their effect on Newtonian noise has not yet been systematically quantified. In particular, the separate roles of sediment-basement interface geometry, sediment shear-wave velocity, and attenuation remain insufficiently understood. Many previous numerical studies have considered deterministic excitation from individual sources, such as single forces or double-couple sources \cite{atterholt2026grl, tian2025basinanalysis}, whereas the response to a spatially distributed ambient-noise field has not yet been investigated. In addition, most seismological site-response studies focus on surface observations, so the dependence of Newtonian noise on test-mass depth and lateral position within the subsurface remains poorly constrained. 

This paper is organized as follows. Section~\ref{sec:method} summarizes the numerical framework and describes the source configuration, two-layer sediment model, and controlled parameter tests. Section~\ref{sec:results} presents the effects of sediment-basement interface geometry, sediment shear-wave velocity $c_S$, attenuation, and test-mass position on the seismic wavefield and Newtonian-noise response. Section~\ref{sec:conclusions} summarizes the main findings and discusses directions for future work.

\section{Numerical framework and experimental design}
\label{sec:method}

This section describes the numerical experimental design used to investigate the first-order influence of the geometry and material properties of a sediment layer on both the seismic wavefield and the resulting Newtonian noise.

\subsection{Numerical framework}

We use the numerical framework developed in \cite{framework} to simulate the seismic wavefield and the resulting Newtonian noise. Since the framework is already described and validated \cite{framework}, we only summarize the main steps relevant to the present sediment-layer experiments.

The seismic wavefield is computed with Salvus \cite{afanasiev2019salvus} by solving the viscoelastic seismic wave equation,

\begin{equation} 
\rho(\vec{x}) \, 
\frac{\partial^2 \vec{\xi}(\vec{x},t)}{\partial t^2} 
= 
\vec{\nabla} \cdot \boldsymbol{\sigma}(\vec{x},t) 
+ 
\vec{f}(\vec{x},t), 
\label{eq:seismic} 
\end{equation}

where $\rho$ is the density, $\vec{\xi}$ is the displacement vector, $\boldsymbol{\sigma}$ is the viscoelastic stress tensor, and $\vec{f}$ represents the external force density. The material response is controlled by the wave velocities $c_P$ and $c_S$, density $\rho$, and quality factors $Q_P$ and $Q_S$. 
Because the framework described in
\cite{framework} assumes an elastic medium, we briefly describe here
how intrinsic attenuation is incorporated. For a linear isotropic viscoelastic medium, the infinitesimal strain tensor is:

\begin{equation}
\varepsilon_{ij}(\vec{x},t)
=
\frac{1}{2}
\left(
\frac{\partial \xi_i}{\partial x_j}
+
\frac{\partial \xi_j}{\partial x_i}
\right),
\label{eq:strain}
\end{equation}

and the stress-strain relation is

\begin{equation}
\sigma_{ij}(\vec{x},t)
=
\int_{0}^{t}
\left[
\lambda(\vec{x},t-\tau)\,\delta_{ij}\,
\dot{\varepsilon}_{kk}(\vec{x},\tau)
+
2\mu(\vec{x},t-\tau)\,
\dot{\varepsilon}_{ij}(\vec{x},\tau)
\right]
\,\mathrm{d}\tau,
\label{eq:stress_strain}
\end{equation}

where $\lambda$ and $\mu$ are the Lam\'e relaxation functions,
$\delta_{ij}$ is the Kronecker delta, and a dot denotes a time
derivative. The relaxation functions are parameterized to reproduce
the prescribed P- and S-wave velocities, $c_P$ and $c_S$, and quality
factors, $Q_P$ and $Q_S$, over the simulated frequency band.

The simulated displacement field is then converted into a density perturbation using the linearized continuity equation, 

\begin{equation} 
\delta\rho(\vec{x},t) = -\vec{\nabla}\cdot \left[ \rho(\vec{x})\vec{\xi}(\vec{x},t) \right], 
\label{eq:NN_continuity_eq} 
\end{equation} 

where $\rho(\vec{x})$ denotes the stationary background density. This formulation accounts for density perturbations caused by both volumetric deformation and displacement across material interfaces. The resulting density perturbation is used to calculate the time-dependent gravitational perturbation at the test-mass position, from which the Newtonian-noise force is obtained. The 3-D volume integral is evaluated from the 2-D model by assuming translational invariance in the out-of-plane direction.

\subsection{Model and source configuration}

\begin{figure}[htbp]
    \centering
    \includegraphics[width=0.95\linewidth]{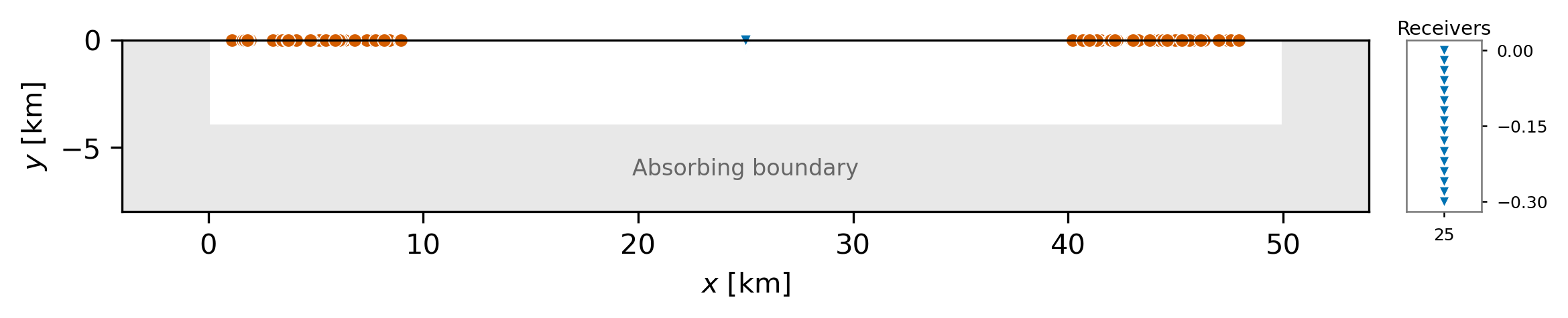}
    \caption{Model geometry and source-receiver configuration. One hundred vertically directed point-force noise sources are distributed along the free surface (red dots), with 50 sources on either side of the receiver array. Sixteen receivers (blue triangle) form a vertical array at $x=25~\mathrm{km}$, extending from the surface to a depth of $300~\mathrm{m}$ at $20~\mathrm{m}$ intervals. The shaded regions denote the absorbing layers.}
    \label{fig:model_geometry}
\end{figure}

We use 2-D simulations to investigate the first-order effects of sediment structure on seismic wavefields and Newtonian noise. The physical model region covers a region of $50~\mathrm{km} \times 4~\mathrm{km}$. To minimize artificial reflections from the model
boundaries, $4~\mathrm{km}$-thick absorbing layers are added along the
left, right, and bottom sides, resulting in a total computational domain
of $58~\mathrm{km}\times8~\mathrm{km}$ (Fig.~\ref{fig:model_geometry})..

A total of 100 random noise sources are distributed along the surface on the left and right sides of the seismic array, with each source represented by an upward vertical single force (Fig.~\ref{fig:model_geometry}). This source geometry illuminates the receiver region from both sides and reduces directional bias in the simulated wavefield. The use of vertical single-force sources provides a simplified representation of surface-generated seismic noise.

\begin{figure}[htbp]
    \centering
    \includegraphics[width=0.9\linewidth]{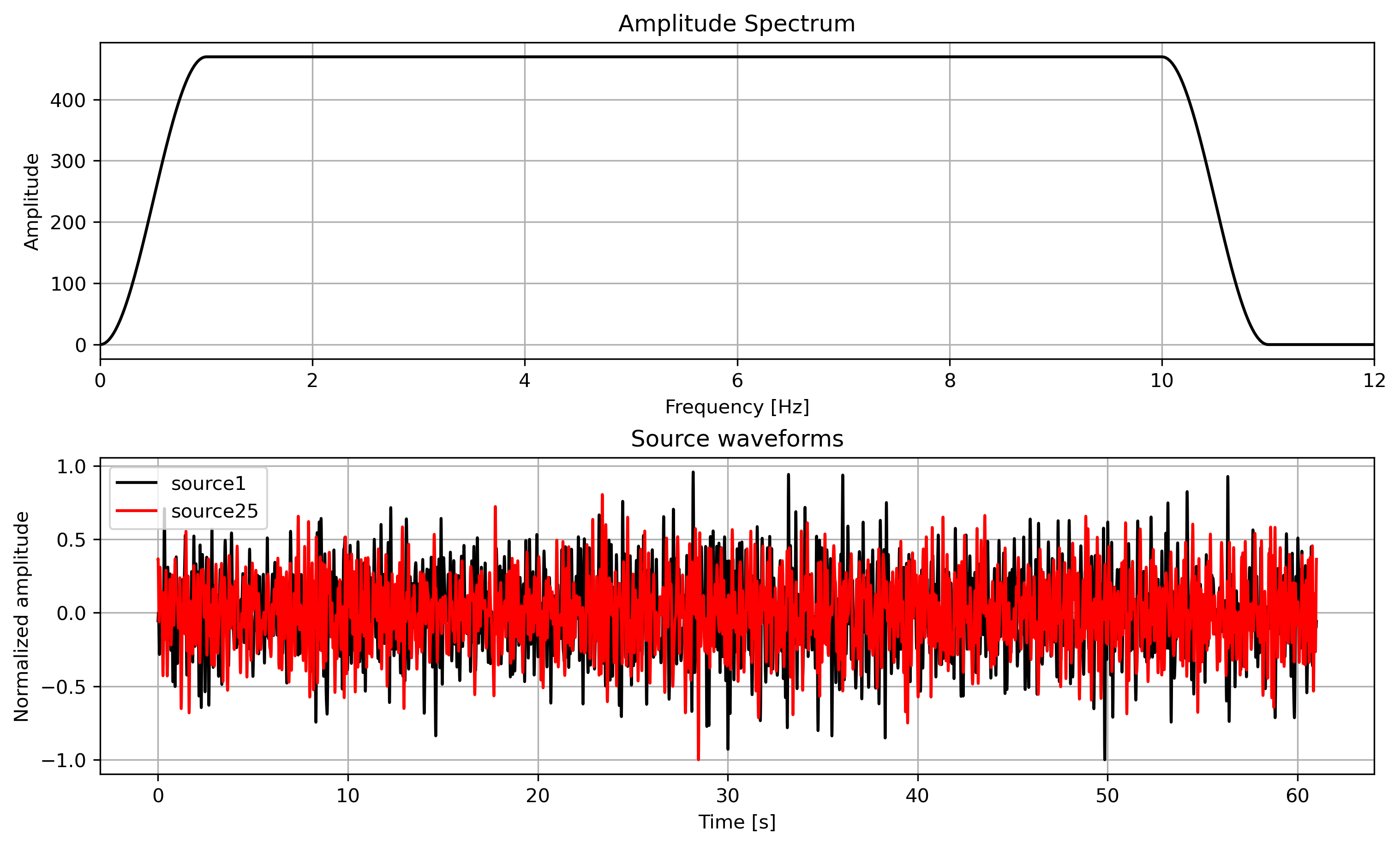}
    \caption{Source spectrum and example source waveforms. The top panel shows the prescribed amplitude spectrum. The bottom panel shows example time functions.}
    \label{fig:source}
\end{figure}

Each source is assigned an independent random-noise time function. The source spectrum is approximately flat between 1 and 10 Hz (Fig.~\ref{fig:source}), and each source has an independent random phase. This choice avoids artificial phase coherence between sources and approximates a spatially distributed stochastic noise field. All sources are active simultaneously during the simulation to imitate continuous ambient seismic excitation, and the resulting wavefield is simulated for 30 seconds.

The same source configuration is used for all models, so that differences in the results can be attributed to changes in the sediment structure rather than to changes in the sources.

We use a 2-D setup as an efficient first step to isolate the first-order effects of sediment structure. For the model configuration considered here, a single 2-D simulation requires
approximately $5$-$10~\mathrm{min}$ when run in parallel on eight CPU cores of an Apple M4 Pro processor. In contrast, a corresponding 3-D simulation with comparable resolution would contain approximately $10^3$-$10^4$ times as many degrees of freedom as the 2-D model, making systematic parameter tests substantially more computationally expensive. Although a 2-D point source represents an out-of-plane line source in 3-D and therefore differs from a 3-D point source in geometrical spreading and spatial coherence \cite{ForbrigerEtAl2014}, these effects are expected primarily to modify the absolute amplitudes rather than the principal physical mechanisms. Direct comparisons of 2-D and 3-D basin simulations indicate that the principal wave-propagation mechanisms and overall response patterns are retained \cite{MakraChavezGarcia2016}.

\subsection{Controlled sediment-layer parameter tests}

To explore the first-order influence of the sediment layer, we design a set of numerical experiments using a simple two-layer sediment model. This idealized setup is motivated by geological investigations of the ET candidate sites, where near-surface sedimentary cover is an important feature of the site characterization.

The upper layer represents unconsolidated sedimentary cover, such as Quaternary deposits, whereas the lower layer represents the consolidated hard-rock basement, such as granite. Sedimentary layers are commonly characterized by low shear-wave velocity, $c_S$, a high $c_P/c_S$ ratio, and a low shear-wave quality factor, $Q_S$. The reference material parameters are taken from the active-seismic measurements \cite{rietbrock2025lausitz} in the Lausatia and are listed in Table~\ref{tab:sediment_parameter}. The sediment layer is assigned low seismic velocities and strong attenuation, whereas the basement is represented by a faster and less attenuating granite layer.

\begin{table}[htbp] \centering \caption{ Material parameters for the two-layer sediment model. Bold values denote
the primary parameters varied in the sensitivity analyses.} \label{tab:sediment_parameter} 
\begin{tabular}{lccccc} 
\hline 
Layer & $c_P$ (m/s) & $c_S$ (m/s) & $\rho$ (kg/m$^3$) & $Q_P$ & $Q_S$ \\ 
\hline Sediment & 1530 & \textbf{510} & \textbf{1800} & \textbf{80} & \textbf{20} \\ 
Basement & 5590 & 3288 & 2700 & 400 & 200 \\ \hline 
\end{tabular} 
\end{table}

To isolate the effects of sediment $c_S$, sediment $Q_S$, and sediment-basement interface geometry, we design three groups of controlled experiments, as summarized in Table~\ref{tab:parameter_tests}. For the material parameter tests, we vary the relevant parameter values. For the sediment–basement interface geometry, we compare a flat sedimentary layer of constant thickness with a basin-shaped sedimentary structure.

\begin{table}[htbp]
\centering
\caption{Controlled sediment-layer parameter tests.}
\label{tab:parameter_tests}
\begin{tabular}{lll}
\hline
Test group & Values or models compared & Fixed parameters \\
\hline
Interface geometry 
& Flat and basin-shaped interface 
& $c_P$, $c_S$, $\rho$, $Q_P$, $Q_S$ \\

Sediment velocity 
& $c_S \in \{400, 600\}$; $\rho$ derived from $c_S$ \cite{brocher2005empirical}
& Interface, $c_P$, $Q_S$, $Q_P$ \\

Sediment attenuation 
& $Q_S \in \{5, 100\}$; $Q_P$ derived from $Q_S$ \cite{ivanov2014qpqs}
& Interface, $c_S$, $c_P$, $\rho$ \\
\hline
\end{tabular}
\end{table}

\section{Results and discussion}
\label{sec:results}

This section presents the main results from the controlled sediment-layer experiments. We first examine how sediment-basement interface geometry affects resonance coherence and Newtonian-noise amplification. We then isolate the effects of sediment $c_S$ and $Q_S$ on the frequency content and amplitude of the seismic response. The spatial dependence of Newtonian noise is then explored by varying both the depth and lateral position of the test mass. Finally, we make a comparison between the basin-sediment model and the homogeneous model. To compare the models, we examine vertical-displacement wavefields, displacement amplitude spectral density (ASD), ASD ratios, and Newtonian-noise strain ASD.

\subsection{Effect of sediment-basement interface geometry on Newtonian noise}

\begin{figure}[htbp]
    \centering
    \includegraphics[width=0.95\linewidth]{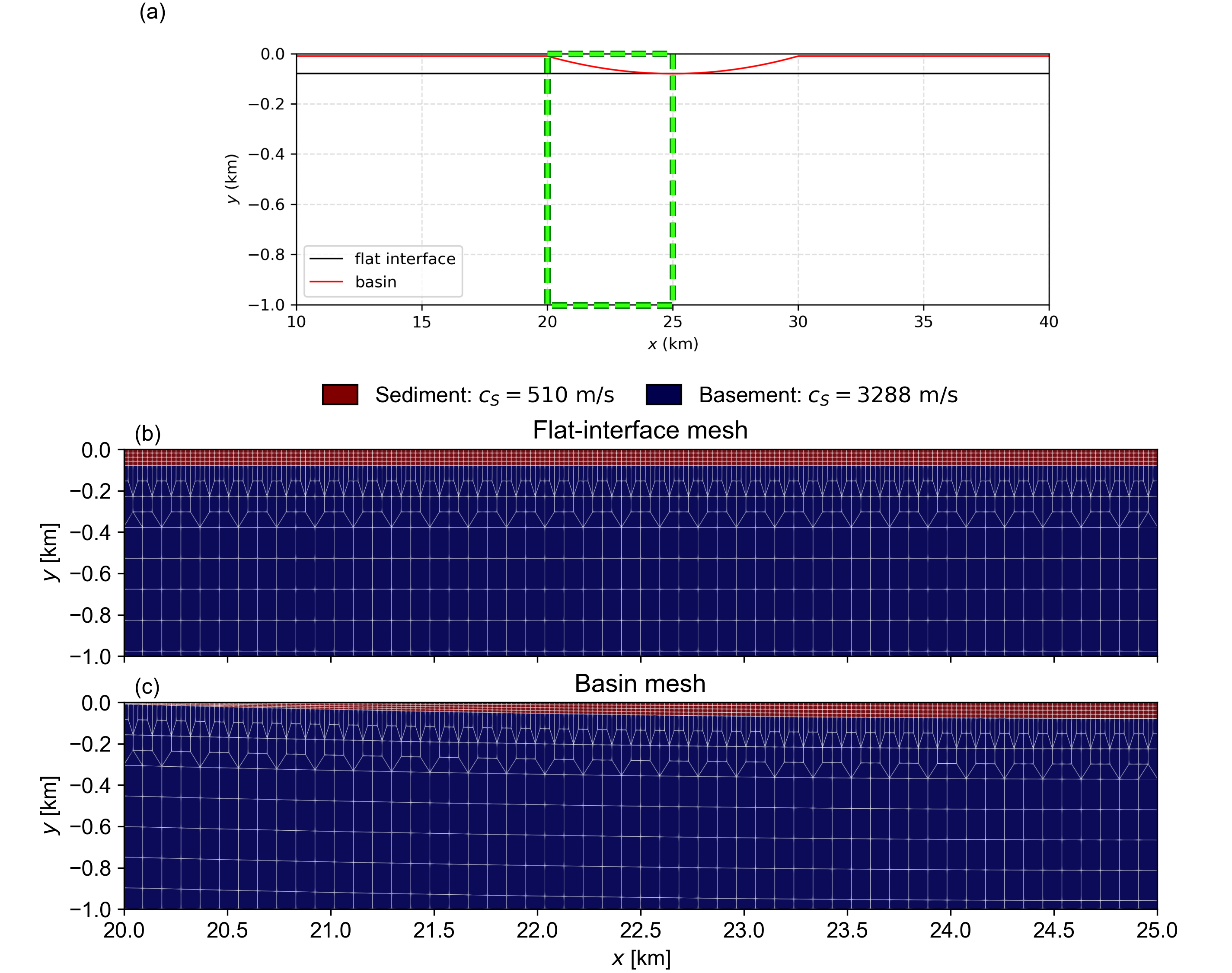}
    \caption{
    Interface geometry and corresponding meshes.
    (a) Comparison of the flat and basin-shaped sediment-basement interfaces. For visibility, the vertical scale is exaggerated by a factor of 10 relative to the horizontal scale. This exaggeration affects only the plotted aspect ratio; all interface depths and sediment thicknesses are shown at their actual values. The green dashed rectangle marks the region shown in panels (b) and (c), spanning \(x=20\text{-}25~\mathrm{km}\) and \(y=-1\text{-}0~\mathrm{km}\).
    (b) Mesh for the flat-interface model.
    (c) Mesh for the basin-shaped model.
    }
    \label{fig:interface_meshes}
\end{figure}

To isolate the effect of sediment-basement interface geometry, we compare a flat-interface model and a basin-shaped model that differ only in the shape of the interface. The interface geometries are shown in Fig.~\ref{fig:interface_meshes}(a). 
The flat-interface model has a constant sediment thickness of $80~\mathrm{m}$, whereas the basin-shaped model has a laterally varying sediment thickness between $10~\mathrm{m}$ and $80~\mathrm{m}$ over the horizontal range $x=20$-$30~\mathrm{km}$. The basin interface is represented by a circular arc passing through
$(20~\mathrm{km},-10~\mathrm{m})$, $(25~\mathrm{km},-80~\mathrm{m})$,
and $(30~\mathrm{km},-10~\mathrm{m})$. The corresponding mesh structures are shown for the zoomed region $x=20$-$25~\mathrm{km}$ and $y=-1$-$0~\mathrm{km}$ in Figs.~\ref{fig:interface_meshes}(b) and \ref{fig:interface_meshes}(c). The meshes are locally refined near the sediment-basement interface to better resolve the interface geometry.

\begin{figure}[htbp]
    \centering
    \includegraphics[width=1.0\linewidth]{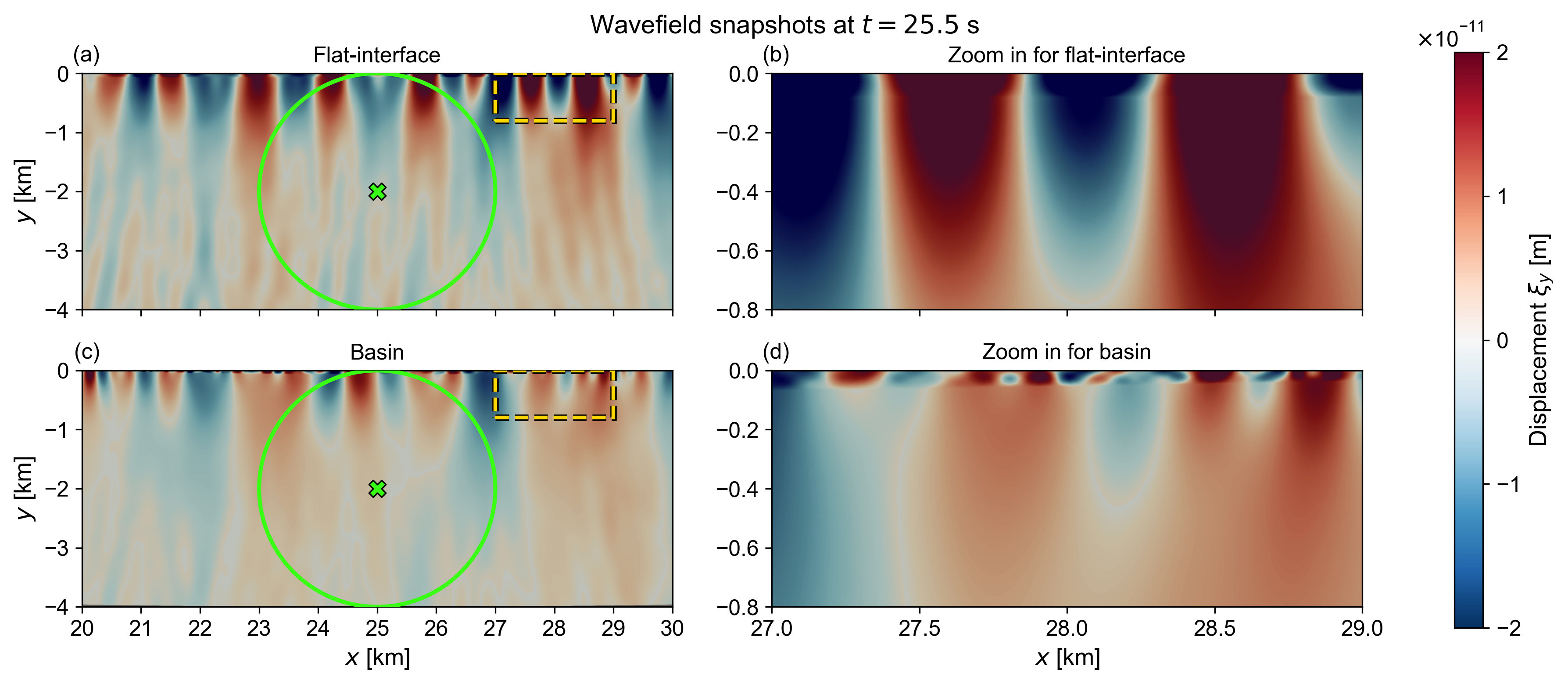}
    \caption{Wavefield snapshots of vertical displacement at $t=25.5\,\mathrm{s}$ for the flat-interface (a,b) and basin (c,d) models. Panels (b) and (d) enlarge the dashed regions in panels (a) and (c), respectively. The green crosses mark the test mass at $(25,-2)\,\mathrm{km}$, while the green circles indicate the integration regions with a radius of $2\,\mathrm{km}$.}
    \label{fig:wavefield_snapshots}
\end{figure}

We first compare the seismic wavefields produced by the two interface geometries. Figure~\ref{fig:wavefield_snapshots} shows snapshots of the vertical displacement for the flat-interface and basin-shaped models at $t=25.5~\mathrm{s}$. The plotted region covers the central $10~\mathrm{km}$ of the model.

For the flat-interface model, the wavefield exhibits broad, laterally coherent regions of alternating polarity as shown in Fig.~\ref{fig:wavefield_snapshots}(a) and (b). This pattern indicates repeated reflections in the low-velocity layer as the seismic wave is trapped due to the strong impedance contrast. Coherent wavefronts extend beneath the sediment layer, indicating transmission across the sediment-basement interface. 

For the basin-shaped model, seismic energy is also concentrated within the sediment layer, owing to the strong impedance contrast. However, the wavefield is less laterally coherent than in the flat-interface case, as shown in Fig.~\ref{fig:wavefield_snapshots}(c,d). The laterally varying sediment thickness disrupts the uniform resonance condition and produces a more spatially scattered wavefield. This suggests that basin geometry can weaken the coherent resonance observed in the flat-interface model.

\begin{figure}[htbp]
    \centering
    \includegraphics[width=0.6\linewidth]{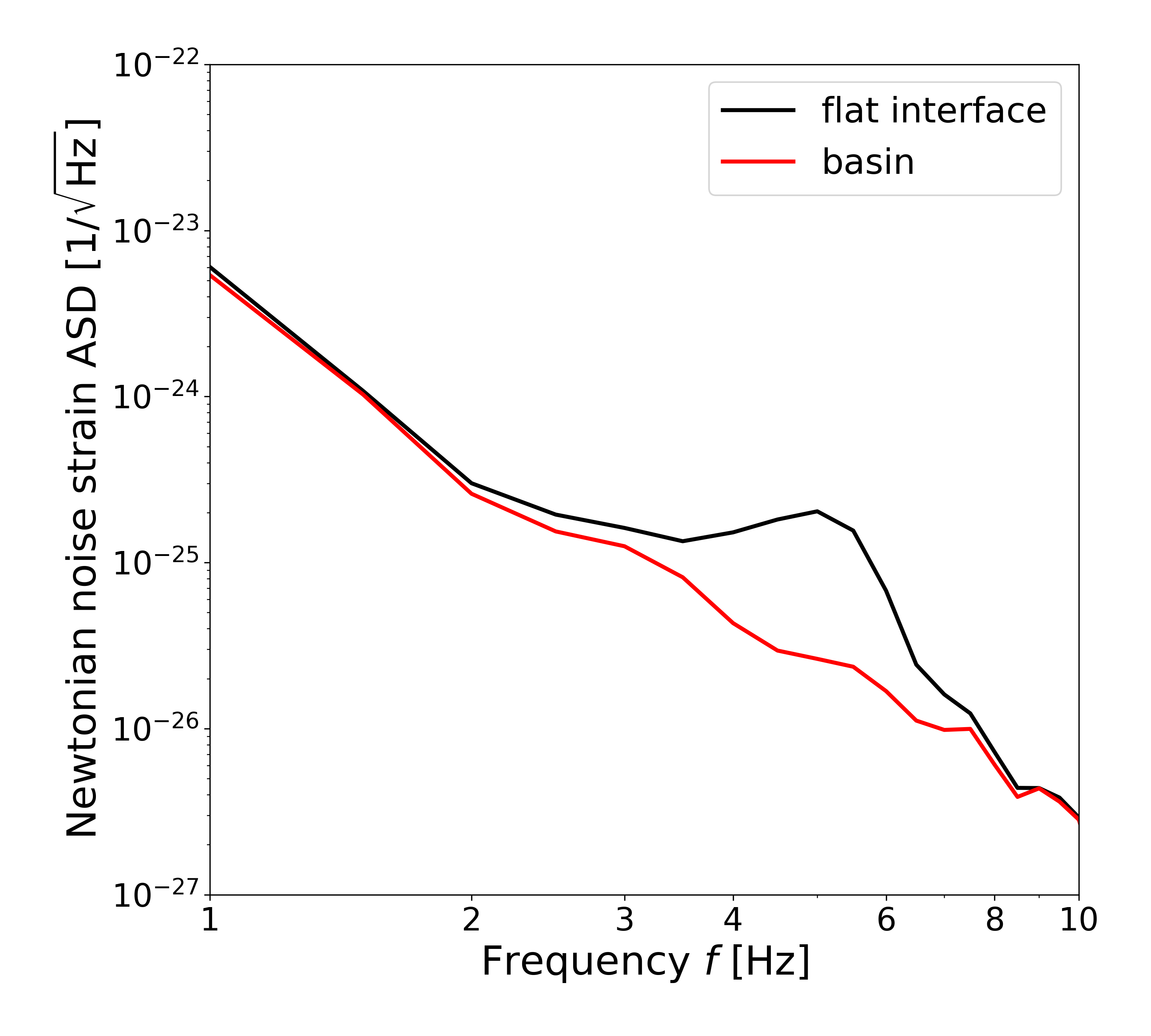}
    \caption{Newtonian-noise strain ASDs at the test-mass position $(25,-2)~\mathrm{km}$ for the flat-interface and basin-shaped models.
    }
    \label{fig:NN_flat_basin}
\end{figure}

We then place a test mass at $(25,-2)~\mathrm{km}$ in both models, as indicated in Figs.~\ref{fig:wavefield_snapshots}(a) and \ref{fig:wavefield_snapshots}(c), and compare the corresponding Newtonian-noise strain ASDs. As shown in Fig.~\ref{fig:NN_flat_basin}, the two models exhibit similar amplitudes at low frequencies, whereas clear differences emerge between approximately $3$ and $7~\mathrm{Hz}$. In particular, the flat-interface model shows a pronounced spectral enhancement around $5~\mathrm{Hz}$, while the basin-shaped model displays a smoother response with lower amplitudes over this frequency range. This suggests that a constant sediment thickness favors stronger and more coherent resonance than a laterally varying basin geometry.

\subsection{Effect of sediment shear-wave velocity}

To investigate the influence of sediment $c_S$, we compare basin-shaped models with sediment shear-wave velocities of $400~\mathrm{m/s}$ and $600~\mathrm{m/s}$. The resulting seismic responses are evaluated at the receiver locations shown in Fig.~\ref{fig:model_geometry}, where the sediment thickness is $80~\mathrm{m}$.

\begin{figure}[htbp]
    \centering
    \includegraphics[width=0.95\linewidth]{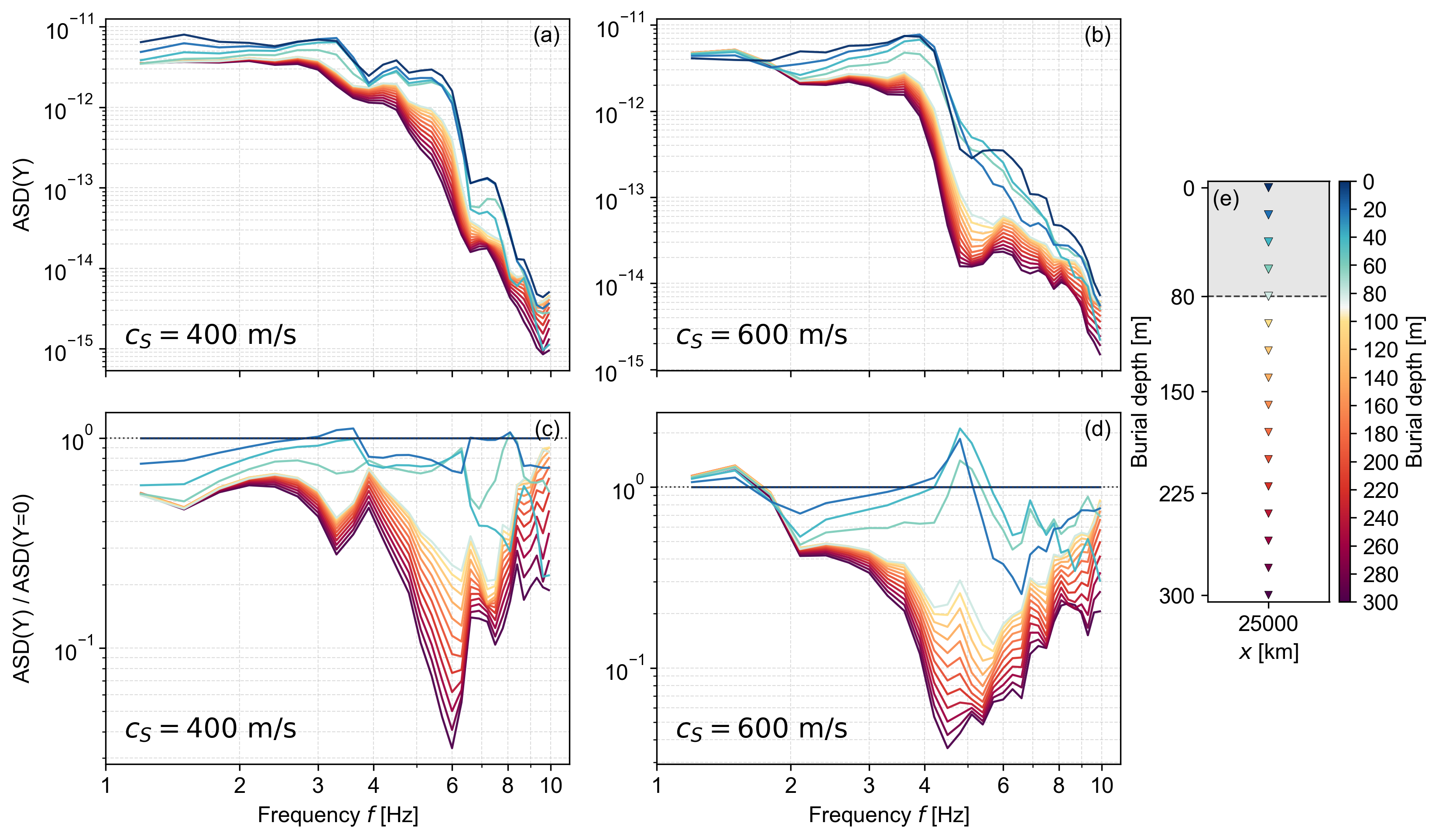}
    \caption{Influence of sediment shear-wave velocity on vertical-displacement ASDs at $x=25~\mathrm{km}$. Panels (a) and (b) show the ASDs for $c_S=400~\mathrm{m/s}$ and $c_S=600~\mathrm{m/s}$, respectively, while panels (c) and (d) show the corresponding ASD ratios relative to the surface station. Panel (e) shows the receiver distribution by burial depth; the dashed line marks the sediment-basement interface.}
    \label{fig:ASD_Vs_compare}
\end{figure}

As shown in Figs.~\ref{fig:ASD_Vs_compare}(a) and (b), both models exhibit a clear contrast between receivers located above and below the sediment-basement interface. The receivers within the sediment layer generally show higher ASD amplitudes than those in the underlying basement, indicating enhanced seismic energy within the low-velocity layer. This behavior is consistent with wave trapping associated with the strong impedance contrast between the sediment and basement.

The corresponding ASD ratios in Figs.~\ref{fig:ASD_Vs_compare}(c) and (d) further highlight the depth dependence of the seismic response. Each curve is normalized by the ASD at the surface receiver.  Receivers in the basement generally show smaller ratios than those within the sediment layer, indicating that the amplified near-surface motion is associated with reduced transmission of seismic energy into the underlying basement.

The two models also show distinct frequency-dependent behavior. For $c_S=400~\mathrm{m/s}$, the lower sediment velocity produces a stronger contrast with the basement and generally smaller subsurface-to-surface ASD ratios over the dominant response band (see Figs.~\ref{fig:ASD_Vs_compare}(a) and \ref{fig:ASD_Vs_compare}(c)). For $c_S=600~\mathrm{m/s}$, The broad spectral peak shifts from approximately $3$-$3.5~\mathrm{Hz}$ to around $4~\mathrm{Hz}$ (see Fig.~\ref{fig:ASD_Vs_compare}(b)). This shift is consistent with the expected increase in characteristic resonance frequency with sediment shear-wave velocity for a fixed layer thickness. The comparison indicates that sediment $c_S$ controls both the frequency range of the seismic response and the strength of wave trapping within the sediment layer.

\subsection{Effect of sediment attenuation}

Near-surface sedimentary cover at candidate ET sites can include weakly consolidated or unconsolidated materials, which are generally associated with strong seismic attenuation and low quality factors. To investigate the influence of sediment attenuation, we conduct two controlled tests with $Q_S=5$ and $Q_S=100$ using the basin-shaped sediment model, representing strongly and weakly attenuating sediment, respectively. 

\begin{figure}[htbp]
    \centering
    \includegraphics[width=0.95\linewidth]{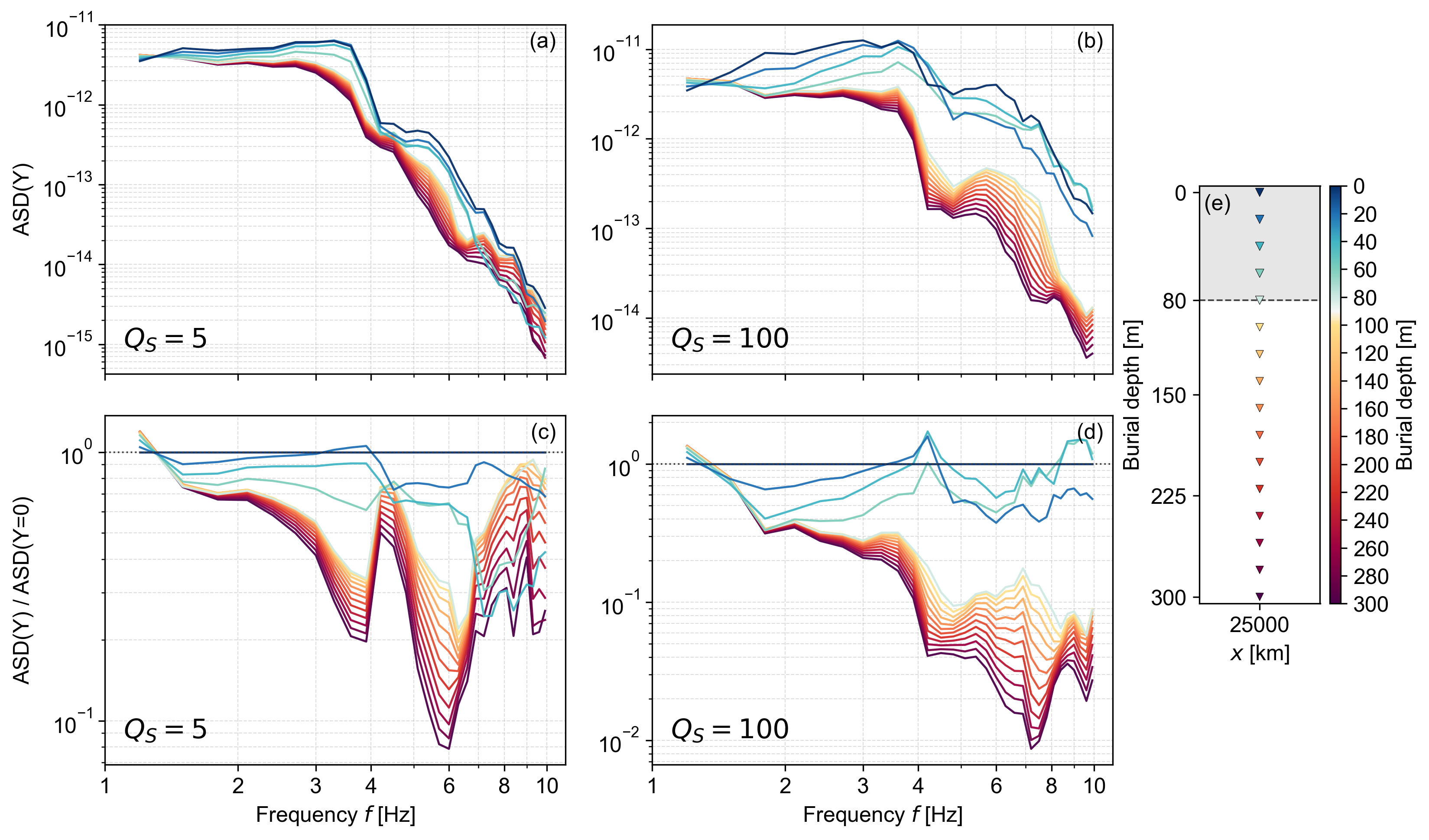}
    \caption{Influence of sediment attenuation on vertical-displacement ASDs at $x=25~\mathrm{km}$. Panels (a) and (b) show the ASDs for $Q_S=5$ and $Q_S=100$, respectively, while panels (c) and (d) show the corresponding ASD ratios relative to the surface station. Panel (e) shows the receiver distribution by burial depth; the dashed line marks the sediment-basement interface.}
    \label{fig:ASD_Qs_compare}
\end{figure}

As shown in Figs.~\ref{fig:ASD_Qs_compare}(a) and (b), the two models exhibit markedly different amplitude responses. For the strongly attenuating case with $Q_S=5$, the ASDs decrease relatively smoothly with frequency, and the separation between near-surface receivers in the sediment layer and deeper receivers in the basement is relatively limited. In contrast, the $Q_S=100$ model shows substantially stronger near-surface ASD amplitudes with a more pronounced separation between receivers above and below the sediment-basement interface. This indicates that stronger attenuation leads to stronger dissipation of seismic energy, whereas weaker attenuation allows trapped energy to persist within the sediment layer, enhancing the frequency-dependent near-surface response.

The corresponding ASD ratios in Figs.~\ref{fig:ASD_Qs_compare}(c) and (d) make this contrast more apparent. For $Q_S=5$, the subsurface-to-surface ratios remain relatively high, with values of approximately $10^{-2}$. For $Q_S=100$, receivers in the basement exhibit much smaller ratios, reaching values of approximately $10^{-4}$ over $6$-$8~\mathrm{Hz}$.

The comparison suggests that sediment attenuation primarily controls the amplitude and persistence of the sediment-layer response. Strong attenuation at low $Q_S$ suppresses coherent amplification and reduces the contrast between surface and subsurface amplitudes, whereas high $Q_S$ allows stronger wave trapping and resonance to develop. The spectral peaks in both models occur at approximately
$3.6$-$3.8~\mathrm{Hz}$ (Fig.~\ref{fig:ASD_Qs_compare}(a) and (b)), indicating that $Q_S$ has a weaker influence on the characteristic frequencies of the response than on its amplitude.

\subsection{Position dependence of Newtonian noise}

To assess the position dependence of Newtonian noise, we evaluate the response at multiple test-mass positions. The analysis considers both vertical variations with burial depth and horizontal variations across the sediment basin. The position-dependent calculations use the basin-shaped reference model with the parameters shown in Table~\ref{tab:sediment_parameter}.

\subsubsection{Depth dependence}

We place test masses at $x=25~\mathrm{km}$ and vary their depths from $y=-20~\mathrm{m}$ to $y=-300~\mathrm{m}$ at intervals of $20~\mathrm{m}$. The corresponding Newtonian-noise strain ASDs are then compared to assess the depth dependence of the response.

\begin{figure}[htbp]
    \centering
    \includegraphics[width=0.9\linewidth]{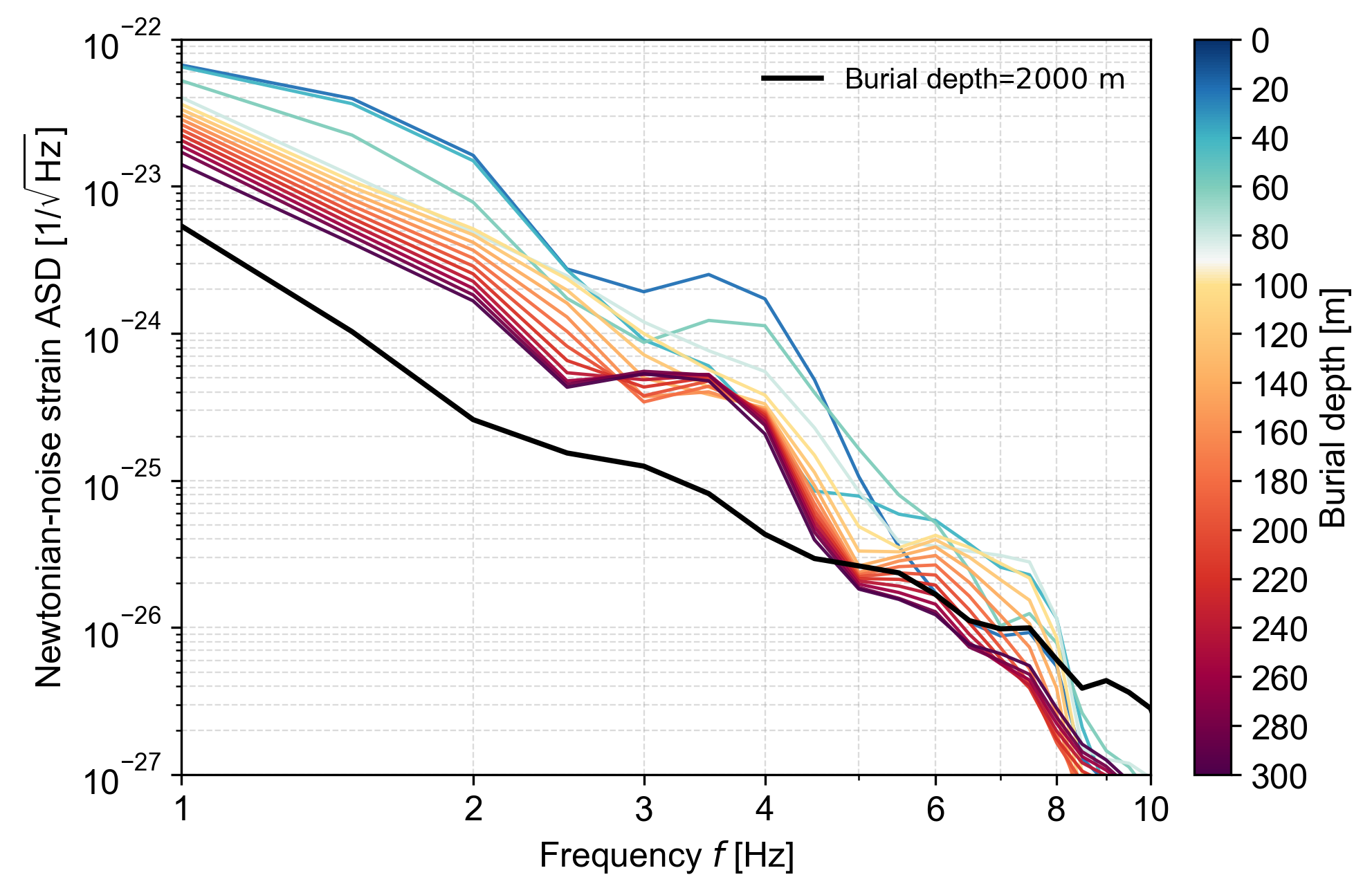}
    \caption{Newtonian-noise strain ASDs at $x=25~\mathrm{km}$ for test-mass depths from $20$ to $300~\mathrm{m}$ below the surface. Colored curves indicate different burial depths, while the solid black curve shows the test mass at $2~\mathrm{km}$ depth.}    
    \label{fig:ASD_depth}
\end{figure}

As shown in Fig.~\ref{fig:ASD_depth}, the Newtonian-noise response exhibits a strong dependence on test-mass depth. The near-surface test masses located within the sediment layer generally show larger amplitudes, particularly at depths of $20~\mathrm{m}$ and $40~\mathrm{m}$. However, the contrast between test masses above and below the sediment-basement interface is less pronounced than that observed in the seismic ASDs (Fig.~\ref{fig:ASD_Vs_compare}). This difference arises because the seismic ASD characterizes the local wavefield at an individual receiver, whereas the Newtonian-noise response is obtained by integrating density perturbations within a $2~\mathrm{km}$ radius around the test mass, thereby incorporating contributions from both the sediment and the basement. 

The depth dependence is most evident below approximately $5~\mathrm{Hz}$. At higher frequencies, the responses at different depths become more similar, because of the stronger attenuation of high-frequency seismic energy in the strongly attenuating sediment layer. In addition, spectral enhancements are observed at approximately
$1.5$-$2~\mathrm{Hz}$, $3$-$4~\mathrm{Hz}$, and around $6~\mathrm{Hz}$-$8~\mathrm{Hz}$
(Fig.~\ref{fig:ASD_depth}), which may be associated with sediment-induced resonance.

The Newtonian-noise ASD for the much deeper reference test mass at $2~\mathrm{km}$ shows substantially lower amplitudes below $5~\mathrm{Hz}$, but becomes comparable to or even exceeds the near-surface responses at higher frequencies (Fig.~\ref{fig:ASD_depth}). This behaviour suggests that sediment-induced
amplification becomes less important with increasing test-mass depth,
particularly at higher frequencies. Because the sources are located far from
the test masses near the centre of the model (Fig.~\ref{fig:model_geometry}),
waves reaching the $2~\mathrm{km}$-deep test mass can be transmitted into the
basement and propagate through this more weakly attenuating medium for much of
their source-receiver path. In contrast, the wavefield at depths of several
hundred meters is more strongly influenced by propagation within the near-surface sediment layer. This difference in propagation
paths may contribute to the relatively strong high-frequency response observed
at $2~\mathrm{km}$ depth. These contrasts demonstrate that a strongly attenuating medium can substantially modify the Newtonian-noise response at different test-mass depths.

\subsubsection{Horizontal dependence}

\begin{figure}[htbp]
    \centering
    \includegraphics[width=0.95\linewidth]{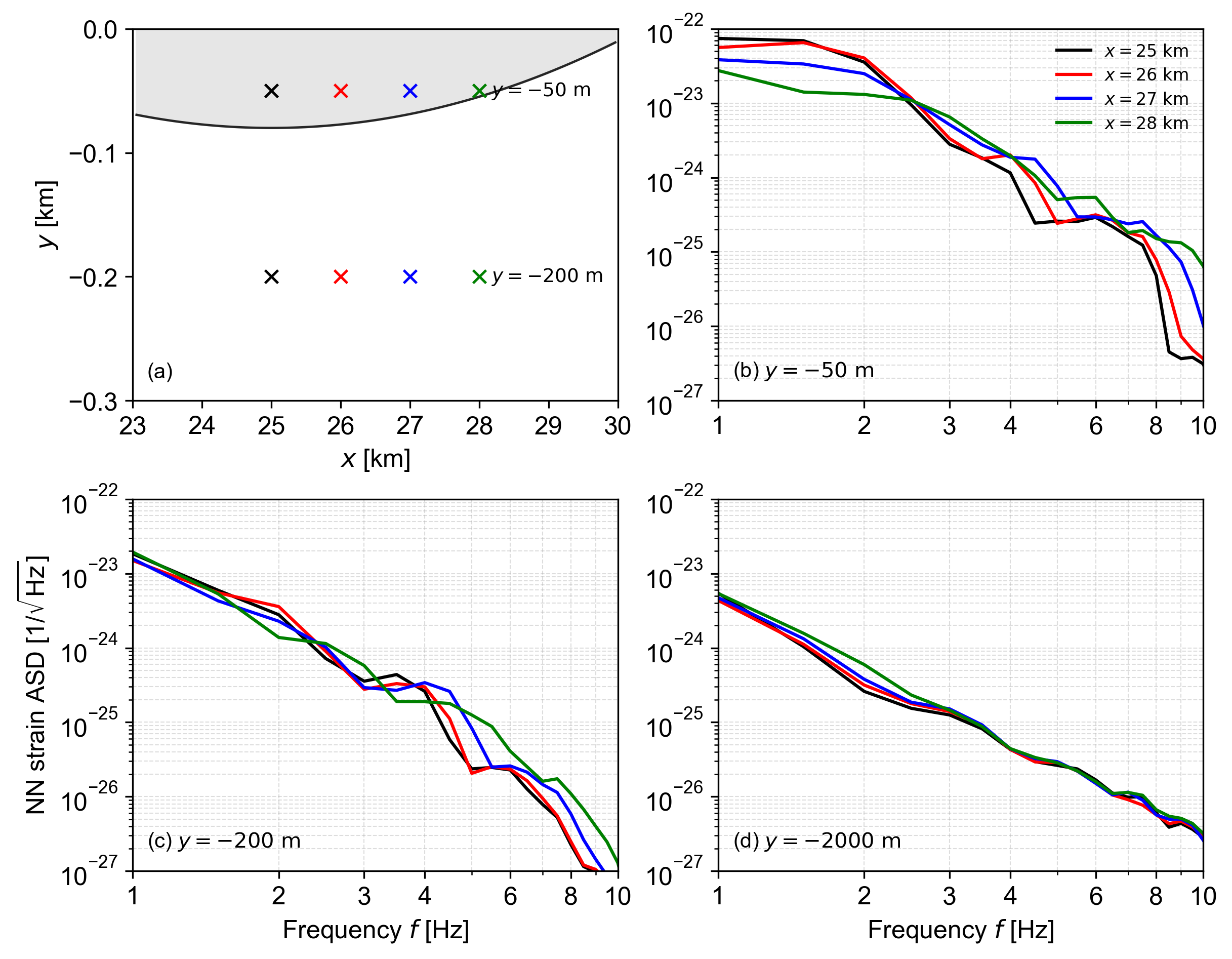}
    \caption{Newtonian-noise strain ASDs for test masses at different horizontal locations and burial depths in the basin-shaped sediment model. Panel (a) shows the sediment-basement interface and the test-mass positions at $y=-50~\mathrm{m}$ and $y=-200~\mathrm{m}$ for $x=25$-$28~\mathrm{km}$. The test masses at $y=-2000~\mathrm{m}$ are omitted from this panel to keep the near-surface geometry visible. Note that the horizontal and vertical axes are plotted with an aspect ratio of $x:y=20:1$. Panels (b)-(d) compare the strain ASDs at $y=-50~\mathrm{m}$, $y=-200~\mathrm{m}$, and $y=-2000~\mathrm{m}$, respectively.}   
    \label{fig:ASD_horizontal}
\end{figure}

To investigate the horizontal dependence of Newtonian noise, we place test masses at four horizontal locations, $x=25$, $26$, $27$, and $28~\mathrm{km}$. These positions correspond to sediment thicknesses of $80$, $77$, $69$, and $55~\mathrm{m}$, respectively. At each horizontal location, test masses are placed at burial depths of $50$, $200$, and $2000~\mathrm{m}$ (Fig.~\ref{fig:ASD_horizontal}(a)). These three depths represent positions within the sediment layer, beneath the sediment-basement interface, and deep within the basement.

For the test masses located within the sediment layer (Fig.~\ref{fig:ASD_horizontal}(b)), the Newtonian-noise amplitudes below approximately $2.5~\mathrm{Hz}$ generally decrease as the sediment becomes thinner. Above $2.5~\mathrm{Hz}$, however, this ordering almost reverses. This crossover likely reflects the thickness dependence of the sediment resonance, with thicker sediment favoring lower characteristic frequencies and thinner sediment shifting the resonant response toward higher frequencies. All four spectra exhibit frequency-dependent enhancements between approximately $3.5$ and $6~\mathrm{Hz}$. These features shift toward higher frequencies as the sediment thickness decreases, consistent with the expected inverse relationship between resonance frequency and layer thickness. Above approximately $8~\mathrm{Hz}$, the thinner-sediment locations generally show larger amplitudes. This behavior may reflect reduced attenuation associated with the smaller sediment thickness and stronger coupling to the less attenuating basement.

For the test masses at $200~\mathrm{m}$ (Fig.~\ref{fig:ASD_horizontal}(c)), the lateral variations remain visible but are weaker than within the sediment layer. Spectral enhancement between $3$ and $6~\mathrm{Hz}$ shows a similar tendency with test masses at $50~\mathrm{m}$ depth, with decreasing sediment thickness shifting the enhancement toward higher frequencies. Above approximately $6~\mathrm{Hz}$, the locations with thinner sediment generally retain larger amplitudes, indicating that the effects of lateral sediment-thickness variations persist below the interface. 

At a depth of $2~\mathrm{km}$ (Fig.~\ref{fig:ASD_horizontal}(d)), the four spectra are nearly identical over most of the frequency range. This indicates that the deep Newtonian-noise response is relatively insensitive to lateral variations in near-surface sediment thickness.

Overall, the horizontal variation of Newtonian noise is strongest for test masses located within or just below the sediment layer. Changes in sediment thickness shift the main spectral features and modify their amplitudes, with thinner sediment generally moving the response toward higher frequencies. These lateral differences become much weaker with depth and are nearly absent at $2~\mathrm{km}$.

\subsection{Overall influence of the sediment layer}

Having examined the separate effects of interface geometry, material properties and test mass positions, we finally evaluate the overall influence of the sediment layer. We compare the basin-shaped sediment model with a homogeneous reference model that contains only the basement material (see Table \ref{tab:sediment_parameter}). The two models use identical source configurations and boundary conditions.

\begin{figure}[htbp]
    \centering
    \includegraphics[width=0.9\linewidth]{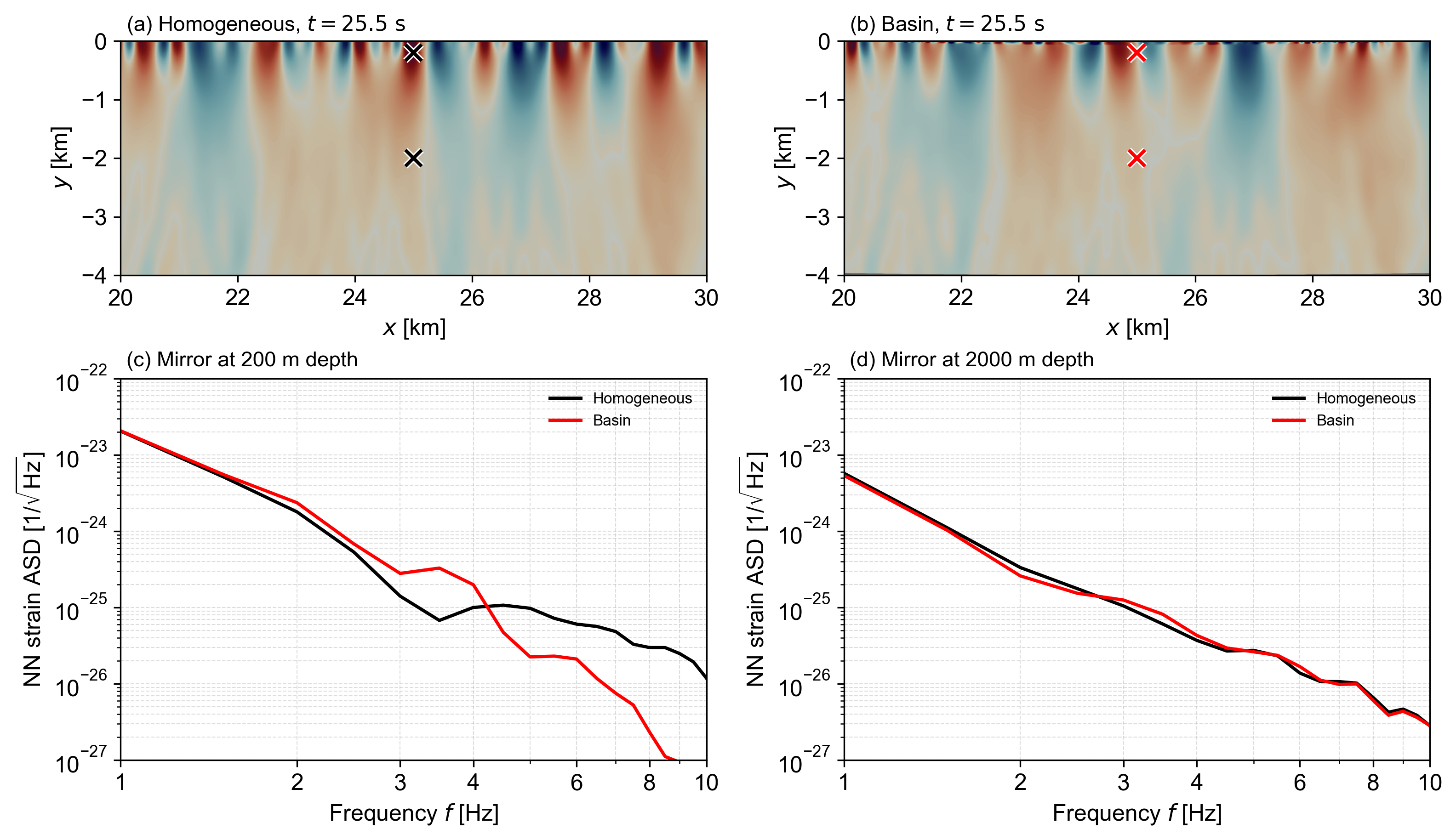}
    \caption{Comparison between the homogeneous and basin-shaped models. Panels (a) and (b) show vertical-displacement wavefield snapshots at $t=25.5~\mathrm{s}$ for the homogeneous and basin-shaped models, respectively. The black and red crosses mark the test-mass positions in homogeneous model and basin-shaped model. Panels (c) and (d) compare the corresponding Newtonian-noise strain ASDs at depths of $200~\mathrm{m}$ and $2000~\mathrm{m}$, respectively.}
    \label{fig:basin_homo}
\end{figure}

Figures~\ref{fig:basin_homo}(a) and \ref{fig:basin_homo}(b) compare snapshots of the vertical-displacement wavefield at $t=25.5~\mathrm{s}$. The basin-shaped model shows a modified and more spatially variable near-surface wavefield (See Fig.~\ref{fig:wavefield_snapshots} (d) for a zoomed-in view), reflecting wave trapping, attenuation, and interaction with the sediment-basement interface. The differences become less pronounced with depth, where the large-scale wavefield patterns in the two models remain broadly similar.

At a test-mass depth of $200~\mathrm{m}$ (Fig.~\ref{fig:basin_homo}(c)), the two models show comparable Newtonian-noise amplitudes below approximately $2.5~\mathrm{Hz}$. The basin-shaped model produces a clear spectral enhancement between approximately $3$ and $4~\mathrm{Hz}$ due to the resonance, followed by lower amplitudes than the homogeneous model at higher frequencies. At a depth of $2~\mathrm{km}$ (Fig.~\ref{fig:basin_homo}(d)), the two spectra are nearly identical over most of the frequency range. This indicates that the influence of the near-surface sediment layer becomes weak for deeply buried test masses. 

Overall, the basin-shaped sediment layer produces comparable Newtonian noise below its characteristic response, enhanced amplitudes near the sediment-related spectral band, and reduced amplitudes at higher frequencies for near-surface test-mass positions.

\section{Conclusions}
\label{sec:conclusions}

We used 2-D viscoelastic simulations to investigate the first-order influence of a near-surface low-velocity sediment layer on the seismic wavefield and the resulting Newtonian noise. Controlled experiments are designed to examine the effects of sediment-basement interface geometry, sediment shear-wave velocity, attenuation, and test-mass position under a random source configuration. 

The simulations show that the sediment layer concentrates seismic energy near the surface and introduces strongly frequency-dependent responses through wave trapping, reverberation, and attenuation. A flat interface with constant sediment thickness produces a more laterally coherent wavefield and a sharper Newtonian-noise spectral
enhancement than the basin-shaped interface. Lateral variations in sediment thickness therefore reduce the coherence of the layered response and broaden the corresponding Newtonian-noise spectrum.

Sediment material properties control different aspects of this response. Lower sediment shear-wave velocity $c_S$ shifts the characteristic spectral features toward lower frequencies and increases the contrast between the sediment and the underlying basement. Sediment attenuation primarily controls the amplitude and persistence of the trapped
wavefield. Low shear-wave quality factors $Q_S$ dissipates seismic energy more strongly, whereas high $Q_S$ preserves stronger near-surface amplification, while producing smaller changes in the frequencies.

The resulting Newtonian noise is strongly position dependent. Near-surface test masses are more sensitive to sediment-related spectral enhancements, while the influence of lateral sediment-thickness variations decreases with burial depth. The depth dependence is not
monotonic at all frequencies, demonstrating that burial depth alone is not sufficient to predict the Newtonian-noise level. Instead, the local sediment thickness, velocity, attenuation, and test-mass position must be considered together.

These results demonstrate that near-surface sedimentary structure should be included explicitly in site-specific Newtonian-noise prediction for the Einstein Telescope. Future work will incorporate measured seismic spectra, 3-D geological structure, topography, caverns, and uncertainty in the sediment properties.

\section*{Availability of Code}
The code used to produce the results of this paper is public at \url{https://github.com/YaoShi0410/Sediment-seismic-newtonian-noise.git}.

\section*{Acknowledgements}
This research was supported by the German Federal Ministry of Research, Technology and Space (BMFTR) via project 05A2023 under grant number 05A23PA1. Shi Yao was partly funded through the STARK Program (grant number 46SK0336X) by the Federal Ministry of Economic Affairs and Energy (BMWE).
\printbibliography
\end{document}